\documentclass[12pt,preprint]{aastex}
\usepackage{psfig,natbib}

\usepackage{palatcm}

\def\la{\mathrel{\hbox{\rlap{\hbox{\lower4pt\hbox{$\sim$}}}\hbox{$<$}}}}
\def\ga{\mathrel{\hbox{\rlap{\hbox{\lower4pt\hbox{$\sim$}}}\hbox{$>$}}}}

\newcommand{\be}{\begin{eqnarray}}
\newcommand{\ee}{\end{eqnarray}}

\newcommand{\msol}{\ifmmode{{\rm M}_\odot}\else{M$_\odot$}\fi}
\newcommand{\foe}{\ifmmode{10^{51}}\else{$10^{51}$}\fi}

\newcommand{\xni}{\ifmmode{{\rm X}_{\rm Ni}}\else{X$_{\rm Ni}$}\fi}
\def\ang{\hbox{\AA}}
\def\Teff{\ifmmode{T_{\rm eff}}\else{\hbox{$T_{\rm eff}$} }\fi}
\def\Rzero{\ifmmode{R_0}\else{\hbox{$R_0$} }\fi}

\def\SP2{{\tt IBM SP2}}

\def\PC2{{\tt PC$^2$}}

\def\logg{\log(g)}
\def\mh{[{\rm M/H}]}

\def\inu{\ifmmode{I_{\nu}}\else{\hbox{$I_{\nu}$} }\fi}
\def\snu{\ifmmode{S_{\nu}}\else{\hbox{$S_{\nu}$} }\fi}
\def\jnu{\ifmmode{J_{\nu}}\else{\hbox{$J_{\nu}$} }\fi}

\def\fep{\ifmmode{{\rm Fe II}}\else\hbox{Fe~II }\fi}

  at 12.0 true pt %scaled \magstep1

\def\phoenix{{\tt PHOENIX}}

\def\water{{H$_2$O}}

\def\phoenix{{\tt PHOENIX}}

\def\water{{H$_2$O}}

\def\b{\beta}

\def\rout{\ifmmode{r_{\rm out}}\else\hbox{$r_{\rm out}$}\fi}
\def\tmax{\ifmmode{\tau_{\rm max}}\else\hbox{$\tau_{\rm max}$}\fi}
\def\tstd{\ifmmode{\tau_{\rm std}}\else\hbox{$\tau_{\rm std}$}\fi}
\def\vmax{\ifmmode{v_{\rm max}}\else\hbox{$v_{\rm max}$}\fi}
\def\muE{\ifmmode{\mu_{\rm E}}\else\hbox{$\mu_{\rm E}$}\fi} 
\def\pE{\ifmmode{p_{\rm E}}\else\hbox{$p_{\rm E}$}\fi} 
\def\bmax{\ifmmode{\b_{\rm max}}\else\hbox{$\b_{\rm max}$}\fi}

\def\ang{\hbox{\AA}}

\def\Msun{\hbox{$\,$M$_\odot$} }

\def\Teff{\hbox{$\,T_{\rm eff}$} }

\def\rout{\hbox{$r_{\rm out}$} }

\def\chistd{\ifmmode{\chi_{\rm std}}\else\hbox{$\chi_{\rm std}$}\fi}

\def\K{\,{\rm K}}
\def\msol{$M_\odot$}
\def\foe{10^{51}}

\def\xni{{\rm X}_{\rm Ni}}

\def\lstar{\ifmmode{\Lambda^*}\else\hbox{$\Lambda^*$}\fi} 
\def\Rop{\ifmmode{[R_{ij}]}\else\hbox{$[R_{ij}]$}\fi}

\def\Rji{\ifmmode{[R_{ji}]}\else\hbox{$[R_{ji}]$}\fi}
\def\Rstar{\ifmmode{[R_{ij}^*]}\else\hbox{$[R_{ij}^*]$}\fi}

\def\Rjistar{\ifmmode{[R_{ji}^*]}\else\hbox{$[R_{ji}^*]$}\fi}
\def\DRji{\ifmmode{[\Delta R_{ji}]}\else\hbox{$[\Delta R_{ji}]$}\fi}
\def\DRij{\ifmmode{[\Delta R_{ij}]}\else\hbox{$[\Delta R_{ij}]$}\fi}

\def\ns{\ifmmode{N_{\rm s}}          % Anzahl der tau-punkte
        \else\hbox{$N_{\rm s}$}\fi}

\def\mat#1{{\bf #1}}     % Macro fr Matrizen
\def\vek#1{{#1}}         % Macro fr Vektoren

\newcount\eqcount
\def
  \nummer{
    \global\advance\eqcount by 1
    (\the\eqcount)
  }

\def
  \numadv{
    \global\advance\eqcount by 1
  }

\def
   \numout#1{
     (\the\eqcount #1)
  }

\def\ivek#1#2{\ifmmode{\vek{I}^{#1}_{#2}}
        \else\hbox{$\vek{I}^{#1}_{#2}$}\fi}

\def\tmat#1#2{\ifmmode{\mat{t}^{#1}_{#2}}
        \else\hbox{$\mat{t}^{#1}_{#2}$}\fi}
\def\rmat#1#2{\ifmmode{\mat{r}^{#1}_{#2}}
        \else\hbox{$\mat{r}^{#1}_{#2}$}\fi}
\def\bvek#1#2{\ifmmode{\beta^{#1}_{#2}}
        \else\hbox{$\beta^{#1}_{#2}$}\fi}

\def\lp{\ifmmode{\lambda^+_\tau}           % lambda +
        \else\hbox{$\lambda^+_\tau$}\fi}
\def\lm{\ifmmode\lambda^-_\tau             % lambda -
        \else\hbox{$\lambda^-_\tau$}\fi}

\chardef\tilt=126
\begin{document}
\bibliographystyle{apj}

\title{Spherically symmetric model atmospheres for low mass pre-Main Sequence
stars with effective temperatures between 2000 and 6800~K}

\author{France Allard}
\affil{C.R.A.L (UML 5574) Ecole Normale Superieure, 69364 Lyon Cedex 7, France\\
E-Mail: \tt fallard@ens-lyon.fr}
\author{Peter H. Hauschildt}
\affil{Dept.\ of Physics and Astronomy \& Center for Simulational Physics, 
University of Georgia, Athens, GA 30602-2451\\
Email: {\tt yeti@hal.physast.uga.edu}}
\author{Andreas Schweitzer}
\affil{Dept.\ of Physics and Astronomy,
University of Georgia, Athens, GA 30602-2451\\
Email: {\tt andy@hal.physast.uga.edu}}

{\em ApJ, in press. Also available at \\
\tt ftp://calvin.physast.uga.edu/pub/preprints
}

\begin{abstract}

We present a grid of spherically symmetric model atmospheres for young pre-MS
stars. This grid spans the parameter range $2000\K \leq \Teff \leq 6800\K$ and
$2.0 \leq \logg \leq 3.5$ for $M=0.1\Msun$, appropriate for low mass stars and
brown dwarfs. A major improvement is the replacement of TiO and \water\ line
lists with the newer line list calculated by the NASA-AMES group, for TiO
(about 175 million lines of 5 isotopes) and for \water\ (about 350 million
lines in 2 isotopes). We provide the model structures, spectra and broad-band
colors in standard filters in electronic form.

\end{abstract}

\section{Introduction}

In a recent paper \cite[][hereafter: NG-giant]{ng-giants} we have
presented a grid of spherically symmetric model atmospheres for stars
with $\logg \leq 3.5$ and in the temperature range $3000\leq \Teff
\leq 6800\K$.  In this paper, we present an update of the NG-giant
grid tuned for young pre-MS (PMS) stars. This PMS grid spans the
parameter range $2000\K \leq \Teff \leq 6800\K$ and $2.0 \leq \logg
\leq 3.5$ for solar abundances.  The models were calculated for very
low mass stars with $M=0.1\Msun$ but in the parameter range considered
the mass of the star changes the synthetic spectra and the atmospheric
structure only marginally. A major difference from our NextGen grids
of model atmospheres \cite[]{ng-hot,ng-giants} is the replacement of
TiO and \water\ line lists with the newer line list calculated by the
NASA-AMES group, \cite{ames-tio} for TiO (about 175 million lines of 5
isotopes) and \cite{ames-water-new} for \water\ (about 350 million
lines in 2 isotopes).

In the next section we give a brief overview over the model
construction and the differences from the NextGen grids. Then we
discuss some results, in particular the effects of the new line lists
and we end with a summary of the paper.

\section{Model calculations}

We have calculated the models presented in this paper using our
multipurpose model atmosphere code \phoenix, version 10.7. Details of
the code and the general input physics setup are discussed in
\cite{ng-hot,ng-giants} and \cite{jcam} and references therein. The
model atmospheres presented here were calculated with the same general
input physics as in NG-giant. However, the change of the line lists
has some impact on the model structure and synthetic spectra (see
below).  Our combined molecular line list includes about 550 million
molecular lines.  These lines are treated with a direct opacity
sampling technique where each line has its individual Voigt (for
strong lines) or Gauss (weak lines) line profile \cite[see][and
references therein for details]{ng-hot}. The number of lines selected
by this procedure depends on the the model parameters. Typically, the
smallest amounts of molecular lines are selected at the cool and hot
ends of the grid, e.g., about 75 million for $\Teff=2000\K$,
$\logg=3.5$ and 80 million for $\Teff=4000\K$, $\logg=2.0$.  The
maximum number of selected molecular lines is about 215 million at
$\Teff=3000\K$, $\logg=2.0$ (all data are for solar abundances).

\section{Results}

We have calculated a grid of solar abundance \cite[table 5 of][]{jaschek95}
model atmospheres and  grids with enhanced metallicities $\mh=+0.3$ as
well as reduced metallicities, $\mh=-0.3$.  The models span a range of
$2000\K\leq \Teff \leq 6800\K$ and $2.0\leq\logg\leq 3.5$. These models were
originally intended to be used for low mass stellar evolution models (Baraffe
et al, in preparation) and thus assume a mass of $0.1\Msun$ for the stars. In
the parameter range (mainly $\logg$) considered here, the synthetic spectra can
also be applied to other masses.  Most of the basic results have been described
in \cite{ng-giants}, so we are concentrating here on comparison with the
NG-giant models.

\subsection{Comparison with NG-giant models}

The largest change from the input physics of the NG-giant models is
due to the different and larger line lists for TiO and water vapor. In
Figures \ref{ng-comp-opt} and \ref{ng-comp-ir} we compare NG-giant
models (dotted lines) to the PMS grid.  The main difference between
these models is the selection of the input line lists, we used the
same version of {\tt PHOENIX} and the same thermodynamical and opacity
data (other than TiO and water vapor) for the calculations.  Both sets
of models were iterated to convergence with their respective setups,
so the differences in the spectra are the results of both direct
opacity changes and changes in the structure of the model atmospheres
due to the different opacities. In the optical spectrum, the result is
generally {\em weaker} TiO bands for the PMS models compared to the
NG-giant models at $\Teff=3000\K$ but slightly {\em stronger} TiO
bands at $\Teff=4000\K$. The situation is slightly different for the
water bands shown in Figure \ref{ng-comp-ir}. The NextGen-type models
show stronger H$_2$O bands with less inter-band opacity than the PMS
models. More importantly, the shape of some water bands are noticeably
different between the two setups. The reason for this behavior is the
changes in the structure of the atmosphere caused by the different
degree of completeness of the water vapor line lists used (see Allard,
Hauschildt \& Schwenke, submitted). For low $\Teff$, the strengthening
of the overall water opacity going from the NG-giant models to the PMS
models produces weaker TiO bands due to changes in the structure of
the atmospheres.  However, in hotter models the water bands are not as
important as the TiO bands and the net effect of the overall slightly
stronger TiO lines produces stronger TiO bands in the optical. These
effects are more pronounced for higher gravities (water opacity is
relatively more important for larger gravities at the same effective
temperature).

In general, the PMS setup applied to M dwarfs produces somewhat better
fits to field stars \cite[]{leg99,lhs1070}, although the water bands
are still not perfectly reproduced by the models
\cite[]{AMESpap}\@. One reason for this are problems with the water
line lists, but other opacity sources (such as dust formation in very
cool models) as well as the treatment of convection in optically thin
layers are additional sources of uncertainty.

\subsection{Effects of metallicity changes}

The effects of metallicity changes on low resolution synthetic spectra
are shown in Figs.~\ref{metal-a}--\ref{metal-d} for models with
$\Teff=3400\K$ and $\Teff=2400\K$ for the extreme values of the
gravity in our grid. For the higher effective temperature,
Figs.~\ref{metal-a} and \ref{metal-b}, the effects of metallicity are
most pronounced in the optical (reduced metallicity causes increased
flux due to decreased TiO opacity) and in region from 1 to $1.5\,\mu$m
(the reaction of the atmosphere causes reduced flux in this region to
compensate for the larger flux in the optical). The water band get
stronger with reduced metallicity due to these redistribution
effects. For lower effective temperatures, Figs.~\ref{metal-c} and
\ref{metal-d}, the effects of metallicity of the spectra are
significantly smaller, in particular for the lower gravity shown in
Fig.~\ref{metal-c}. Here the temperatures are so low that the bands
are saturated and thus will not change much within the range of
metallicities considered here (larger changes will eventually affect
the spectra). Significant changes occur only in localized bands, e.g.,
the metal hydrides and other non-saturated bands such as VO.

\subsection{Formation of radiative and convective zones}

Typically, a cool stellar atmosphere has only one convective zone at
the bottom of the atmosphere whereas the top of the atmosphere is (and
has to be) in radiative equilibrium.  The convective zone at the
bottom of the atmosphere connects to the convective envelope of the
interior of the star.  However, our calculations indicate that the
convective region at the bottom of the atmosphere can be disrupted by
the onset of an isolated radiative zone within specific parameter
ranges.  These ranges are illustrated in Fig. \ref{multi}.  Each
symbol represents a model with multiple convective and radiative
zones.  The plot shows that a continuous and not an arbitrary
parameter range exhibits this behavior.

We investigated the cause of this effect and found it to be due to the
relative strengths of H$^-$ absorption and H$_2$O absorption.  H$^-$
absorption is strongest in the inner part of the atmosphere whereas
H$_2$O absorption is strongest in the outer part. The maximum of the
H$_2$O absorption is in layers of the atmosphere with electron
temperatures of roughly 2500 to 3500~K.  If in this region the H$^-$
absorption is weak enough so that the slope with depth of the overall absorption
coefficient is significantly affected by H$_2$O absorption, an inner
radiative zone forms.  In that case the total absorption coefficient
drops fast enough toward the outer boundary to make the atmosphere
transparent enough to form a radiative zone. As soon as water forms
and the H$_2$O absorption becomes strong enough, the energy is more
efficiently carried by convection until the final radiative zone forms
at the very outside of the atmosphere.

For the hottest models with the lowest log(g) (i.e. the models left
and below the ``multiple zone strip'' in Fig.~\ref{multi}) the single
convective zone at the bottom of the atmosphere is substantially
different from that of the models right above the ``multiple zone
strip''.  For the hot models with low log(g), the water absorption
will never become strong enough to change the slope of the total
absorption and the intermediate radiative zone becomes large enough to
remove the intermediate convective zone.  In the cool models with high
log(g), the water absorption is strong already deep inside the
atmosphere and dominates the slope of the total absorption
coefficient.

This behavior is demonstrated in Fig. \ref{kappaplot} where the most
important continuous absorption coefficients have been plotted against
optical depth.  In the top graph the water absorption changes the
steep slope of the total absorption already in the innermost part.  In
the middle two plots, water forms further out and leaves a steep
enough slope in the absorption coefficient to produce an intermediate
radiative zone. In the graph at the bottom, the water absorption can
no longer change the slope imposed by the H$^-$ absorption (they are
almost parallel in the outer regions) and the energy is transported by
radiation.

\section{Summary and Conclusions}

In this paper we presented a grid of spherically symmetric model
atmospheres for pre-MS stars. The main change with respect to the
NG-giant grid is the use of new TiO and water vapor line lists. In the
parameter range considered for the PMS models the changes in the
structures of the atmospheres compared to similar models with the
NextGen setup are relatively small but the differences of the optical
and IR spectra are noticeable.  We provide the model structures,
spectra and broad-band colors in standard filters through the WWW and
anonymous FTP for general use, see {\tt
http://dilbert.physast.uga.edu/\tilt yeti} or {\tt
ftp://calvin.physast.uga.edu/pub/PMS}.  In a forthcoming paper we will
discuss dust formation for cool dwarfs and giants that incorporate the
opacities used in this paper, and will explore separately the effects
of the present PMS models on evolution tracks for pre-main sequence
stars.

\acknowledgments 

This work was supported in part by the CNRS, INSU and by NSF grant
AST-9720704, NASA ATP grant NAG 5-8425 and LTSA grant NAG 5-3619, as
well as NASA/JPL grant 961582 to the University of Georgia, NASA LTSA
grant NAG5-3435 and NASA EPSCoR grant NCCS-168 to Wichita State
University.  This work was supported in part by the P\^ole
Scientifique de Mod\'elisation Num\'erique at ENS-Lyon.  Some of the
calculations presented in this paper were performed on the CNUSC IBM
SP2, the IBM SP2 and SGI Origin 2000 of the UGA UCNS, on the IBM SP of
the San Diego Supercomputer Center (SDSC), with support from the
National Science Foundation, and on the Cray T3E of the NERSC with
support from the DoE.  We thank all these institutions for a generous
allocation of computer time.

\clearpage

\bibliography{yeti,opacity,mdwarf,radtran,general,opacity-fa,mdwarf-fa}

\clearpage
\section{Figures}

\begin{figure}[b]
\caption[]{\label{ng-comp-opt} Comparison of
the models presented in this paper (full line) to NG-giant models
(dotted line) in the optical spectrum. The resolution has been 
reduced to $5\ang$. The top panel shows models with
$\Teff=3000\K$, the bottom panel for $\Teff=4000\K$, both sets have
$\logg=2.0$ and solar abundances.}
\end{figure}

\begin{figure}[b]
\caption[]{\label{ng-comp-ir} Comparison of
the models presented in this paper (full line) to NG-giant models
(dotted line) in the near IR spectral range. The resolution has been 
reduced to $15\ang$. The top panel shows models with
$\Teff=2000\K$, the middle panel shows $\Teff=2000\K$, and
the bottom panel for $\Teff=4000\K$, all three sets have
$\logg=2.0$ and solar abundances.}
\end{figure}

\begin{figure}[b]
%\plotone{f3.eps}
\caption{
\label{multi}
Models with multiple radiative/convective zones.
Each point marks a model that has multiple radiative/convective zones.
To minimize the number of figures we plotted all metallicities
in one graph and used different symbols for different metallicities
as indicated in the figure.
The log(g) values for sub- and super-solar metallicities have
been slightly shifted to keep the figure legible.
Note the continuous distribution of models with
multiple radiative/convective zones, which is also
continuous is metallicity space.
}
\end{figure}

\begin{figure}[b]
\epsscale{0.4}
%\plotone{f4.eps}
\caption{
\label{kappaplot}
Absorption coefficients versus optical depth at 1.2\micron.
All models have z=-0.3 and log(g)=2.5. The effective temperature
of the models are from top to bottom 2800~K, 3000~K, 3200~K and
3400~K.
The cross hatched regions are the convective zones.
For reference, we indicate the gas pressure and the gas temperature
at the respective optical depth at the top of each figure.
To the right we labeled the most important opacity sources.
}
\end{figure}

\begin{figure}[b]
\caption[]{\label{metal-a} Comparison of
low-resolution (about $50\ang$) synthetic spectra for models with $\Teff=3400\K$, 
$\logg=2.0$ and solar abundances (full line), $\mh=-0.3$ (dotted line)
and $\mh=+0.3$ (dashed line).}
\end{figure}

\begin{figure}[b]
\caption[]{\label{metal-b} Comparison of
low-resolution (about $50\ang$) synthetic spectra for models with $\Teff=3400\K$, 
$\logg=3.5$ and solar abundances (full line), $\mh=-0.3$ (dotted line)
and $\mh=+0.3$ (dashed line).}
\end{figure}

\begin{figure}[b]
\caption[]{\label{metal-c} Comparison of
low-resolution (about $50\ang$) synthetic spectra for models with $\Teff=2400\K$, 
$\logg=2.0$ and solar abundances (full line), $\mh=-0.3$ (dotted line)
and $\mh=+0.3$ (dashed line).}
\end{figure}

\begin{figure}[b]
\caption[]{\label{metal-d} Comparison of
low-resolution (about $50\ang$) synthetic spectra for models with $\Teff=2400\K$, 
$\logg=3.5$ and solar abundances (full line), $\mh=-0.3$ (dotted line)
and $\mh=+0.3$ (dashed line).}
\end{figure}

\end{document}